\documentclass[pra,twocolumn,superscriptaddress,showpacs,amsmath,amstex,amssymb,citeautoscript]{revtex4-2}
\usepackage[colorlinks=true,citecolor=blue,linkcolor=blue,breaklinks=true]{hyperref}

\usepackage{color,graphicx}
\usepackage{bm}
\usepackage{amsmath}
\usepackage{amssymb}
\usepackage{mathrsfs}
\usepackage{times}
\allowdisplaybreaks
 \usepackage{caption}
\usepackage{subcaption}

\usepackage[T1]{fontenc}
\usepackage[utf8]{inputenc}
\usepackage{lipsum}
\usepackage{amsmath}
\usepackage{amssymb}
\usepackage{bbm}
\usepackage{braket}
\usepackage{xcolor}
\usepackage{pifont}
\usepackage[mathscr]{euscript}
\usepackage[shortlabels]{enumitem}
\usepackage[justification=justified]{subcaption}
\usepackage{graphicx}
\usepackage{lipsum}
\allowdisplaybreaks
\usepackage{float}
\usepackage{graphicx}
\usepackage{dsfont}
\usepackage{comment}
\usepackage[colorlinks=true]{hyperref}  
\hypersetup{
    bookmarks=true,         
    unicode=false,          
    pdftoolbar=true,        
    pdfmenubar=true,        
    pdffitwindow=false,     
    pdfstartview={FitH},    
    pdftitle={The Kondo effect in Moir\'e TMD: Heavy fermions or disorder?},    
    pdfauthor={Poduval, Laubscher and Das Sarma},     
    pdfsubject={},   
    pdfcreator={},   
    pdfproducer={}, 
    pdfkeywords={} {} {}, 
    pdfnewwindow=true,      
    colorlinks=true,       
    linkcolor=blue, 
    citecolor=blue,        
    filecolor=magenta,      
    urlcolor=blue           
}

\newcommand{\bk}{\vec{k}}
\newcommand{\bq}{\vec{q}}

\newcommand{\figref}[1]{Fig.~\ref{#1}}

\renewcommand{\approx}{\simeq}

\renewcommand{\vec}[1]{\boldsymbol{#1}}

\definecolor{wrongultramarine}{rgb}{1,0.5,0}

\begin{document}
	
\newcommand {\beq} {\begin{equation}}
\newcommand {\eeq} {\end{equation}}
\newcommand {\bqa} {\begin{eqnarray}}
\newcommand {\eqa} {\end{eqnarray}}
\newcommand {\ca} {\ensuremath{c^\dagger}}
\newcommand {\ba} {\ensuremath{b^\dagger}}
\newcommand {\Ma} {\ensuremath{M^\dagger}}
\newcommand {\psia} {\ensuremath{\psi^\dagger}}
\newcommand {\fbar} {\ensuremath{\bar{f}}}
\newcommand {\psita} {\ensuremath{\tilde{\psi}^\dagger}}
\newcommand{\lp} {\ensuremath{{\lambda '}}}
\newcommand{\A} {\ensuremath{{\bf A}}}
\newcommand{\QQ} {\ensuremath{{\bf Q}}}
\newcommand{\kk} {\ensuremath{{\bf k}}}
\newcommand{\qq} {\ensuremath{{\bf q}}}
\newcommand{\kp} {\ensuremath{{\bf k'}}}
\newcommand{\rr} {\ensuremath{{\bf r}}}
\newcommand{\rp} {\ensuremath{{\bf r'}}}
\newcommand {\ep} {\ensuremath{\epsilon}}
\newcommand{\nbr} {\ensuremath{\langle r r' \rangle}}
\newcommand {\no} {\nonumber}
\newcommand{\up} {\ensuremath{\uparrow}}
\newcommand{\dn} {\ensuremath{\downarrow}}
\newcommand{\rcol} {\textcolor{red}}
\newcommand{\bcol} {\textcolor{blue}}
\newcommand{\lt} {\left}
\newcommand{\rt} {\right}
\newcommand{\dt}{\phantom{\tiny 1}}

\newcommand{\kat}[1]{\textcolor{blue}{#1}}


\title{Apparent Kondo effect in Moir\'e TMD bilayers: Heavy fermions or disorder?}
	\author{Prathyush P. Poduval}
 	\thanks{These authors contributed equally to this work.}
	\affiliation{Condensed Matter Theory Center and Joint Quantum Institute, Department of Physics, University of Maryland, College Park, MD 20742, USA}
	\author{Katharina Laubscher}
 	\thanks{These authors contributed equally to this work.}
	\affiliation{Condensed Matter Theory Center and Joint Quantum Institute, Department of Physics, University of Maryland, College Park, MD 20742, USA}
	\author{Sankar Das Sarma}
	\affiliation{Condensed Matter Theory Center and Joint Quantum Institute, Department of Physics, University of Maryland, College Park, MD 20742, USA}

\date{\today }

\begin{abstract}
A recent work by Zhao \textit{et~al.}~\cite{zhao2022gate} reports the realization of a synthetic Kondo lattice in a gate-tunable Moir\'e TMD bilayer system. The observation of a Kondo lattice is supported by a plateau (or dip, depending on filling) in the temperature dependence of the resistivity $\rho(T)$ around $T^*\sim 40$~K, which is interpreted as the Kondo temperature scale, and an apparent enhancement of carrier mass extracted from the low-temperature resistivity data, indicating the emergence of `heavy fermions'. The latter observation is crucially based on the assumption that the primary resistive scattering mechanism is Umklapp electron-electron scattering in the underlying Fermi liquid. In this work, we analyze the experimental data under the assumption that the primary resistive scattering mechanism is \textit{not} electron-electron scattering, but Coulomb scattering by random quenched charged impurities and phonon scattering. We show that a combination of impurity and phonon scattering is a plausible alternative explanation for the observed resistivity that can describe the key features of the experimental data, even if no Kondo lattice has formed, indicating that further theoretical and experimental work is needed to conclusively verify the formation of a Kondo lattice in Ref.~\cite{zhao2022gate}.
\end{abstract}


\maketitle

\section{Introduction}
The possibility of realizing a synthetic Kondo lattice and heavy fermions in multi-layer Moir\'e materials has attracted significant interest over the past few years~\cite{ramires2021emulating,dalal2021orbitally,vano2021artificial,kumar2022gate,guerci2022chiral,song2022magic,chou2022kondo,hu2023kondo,hu2023strongly,yu2023mesoscale}. Following recent theoretical proposals, an important experimental work~\cite{zhao2022gate} has reported the observation of a Kondo lattice in a hole-doped MoTe$_2$ / WSe$_2$ Moir\'{e} TMD heterobilayer system. Here, the MoTe$_2$ layer is brought into a Mott-insulating state hosting localized magnetic moments, while the WSe$_2$ layer provides essentially free itinerant holes that couple to the localized spins via the Kondo exchange coupling. Based on an assumed interplay of these ingredients, Ref.~\cite{zhao2022gate} infers the realization of a Kondo lattice when the Fermi level is tuned inside the Mott gap of the MoTe$_2$ layer. The experimental characteristic features of the purported Kondo lattice are probed indirectly by temperature-dependent resistivity measurements in Ref.~\cite{zhao2022gate}: First, under the assumption that electron-electron scattering of the underlying Fermi liquid is the main resistive scattering mechanism, the effective quasiparticle mass is extracted by fitting a quadratic function to the low-temperature resistivity-vs-temperature data~\cite{mineev2021electron}. This yields a large value of $m\approx 5-10m_e$ (here $m_e$ is the bare electron mass) in the region of the phase diagram where the Kondo lattice is expected. Indeed, such a mass would be consistent with a Kondo lattice, where the strong Kondo exchange coupling leads to the emergence of quasiparticles with a large effective mass (`heavy fermions')~\cite{coleman2015heavy,coleman2001fermi}. Second, beyond a certain characteristic temperature $T^*\sim 40$~K, the resistivity-vs-temperature curve starts to flatten out or even decrease with temperature. In Ref.~\cite{zhao2022gate}, this characteristic temperature is interpreted as the Kondo temperature.

The Kondo physics interpretation in Ref.~\cite{zhao2022gate} is based entirely on the low-temperature resistivity measurement being interpreted as arising from electron-electron scattering with other resistive scattering mechanisms assumed negligible. Previous studies have, however, shown that even non-Kondo two-dimensional (2D) electron gases may, under suitable conditions, manifest a rich and non-monotonic temperature dependence of the resistivity due to an interplay of different competing scattering mechanisms that become operational in different temperature regimes. In 2D semiconductors, including possibly 2D Moir\'{e} TMDs, screened disorder scattering~\cite{dassarma1986theory,dassarma1991charged,sarma2003low,sarma2004metallicity,das2015screening,ahn2022temperature,ahn2022disorder} typically leads to a resistivity that increases linearly with temperature for $T\ll T_F$, where $T_F$ is the Fermi temperature. At higher temperatures, the system undergoes a quantum-to-classical crossover from the strongly screened quantum regime at $T\ll T_F$ to the classical regime at $T>T_F$, where the resistivity now decreases as $1/T$. At the same time, at high temperatures $T>T_{BG}$, where $T_{BG}$ is the Bloch-Gr\"{u}neisen temperature, an additional linear-in-$T$ contribution to the resistivity due to electron-phonon scattering becomes relevant~\cite{min2012interplay,hwang2019linear}. Depending on the relative magnitudes of $T_{BG}$ and $T_F$, different dependences of the resistivity on temperature are possible arising simply from the interplay of impurity scattering and phonon scattering. In particular, when $T_{BG}>T_F$, a flattening or even a local minimum of the resistivity may arise in the intermediate temperature regime ($T\sim T^*$), with the behavior being a linear-in-$T$ increase (from impurity scattering) below $T_F$ and a linear-in-$T$ increase (from phonon scattering) for $T>T_{BG}$. This can give rise to resistivity-vs-temperature behavior manifesting the same qualitative features observed in the resistivity data of Ref.~\cite{zhao2022gate}. Therefore, we believe that an analysis just based on the temperature-dependent resistivity data is not sufficient to conclusively identify the onset of a Kondo lattice phase since the possibility that the resistivity is controlled by impurity and phonon scattering rather than Umklapp electron-electron scattering cannot be ruled out.

To show that an explanation based on disorder and phonon scattering is compatible with the resistivity data reported in Ref.~\cite{zhao2022gate}, we first present empirical fits to the experimental data taking into account the different scattering mechanisms discussed above. We show that the transport signatures reported in Ref.~\cite{zhao2022gate} can be qualitatively explained by a combination of screened disorder scattering (governing the low to intermediate temperature regime) and electron-phonon scattering (dominant in the high-temperature regime). While we also provide the physical parameters extracted from our fits, these should be viewed as crude qualitative estimates at best rather than quantitatively exact numbers since there are considerable uncertainties associated with the experimental data, particularly with regards to the true carrier density, the disorder configuration, and the residual $T=0$ resistivity at various electric fields---in fact, the resistivity data at low temperatures have substantial unexplained variations making a precise quantitative extraction of the underlying parameters impossible~\cite{wenjinprivate}. Based on our analysis, we conclude that a combination of disorder and phonon scattering provides a reasonable explanation for the reported temperature-dependent resistivity data, and that additional experimental signatures are required to conclusively verify the observation of a Kondo lattice. Our work shows that impurity and phonon scattering together can explain the temperature dependent resistivity of Ref.~\cite{zhao2022gate} without resorting to Umklapp electron-electron scattering effects at all.

\section{Model and scattering mechanisms}
Ref.~\cite{zhao2022gate} discusses a MoTe$_2$ / WSe$_2$ Moir\'{e} TMD system with hole doping, where the MoTe$_2$ layer is an insulator while the holes in the WSe$_2$ layer are itinerant. To analyze the resistivity in such a system, we consider transport by 2D itinerant carriers modeled by a parabolic dispersion $\epsilon_{\bk}= \hbar^2k^2/2m$. Here, $m$ is the effective hole mass, {which Ref.~\cite{zhao2022gate} reports to be $m\approx 0.5m_e$,} and $\bk$ is the 2D wave-vector. The system is further characterized by the 2D carrier density $n$, the background lattice dielectric constant $\kappa$ (which we take to be $\kappa=10$), and the total (spin) degeneracy $g=2$. The resistivity is given by the Drude formula 
\begin{align}
    \rho(T) = \frac{m}{ne^2\tau(T)},
\end{align}
 where $\tau(T)$ is the finite-temperature transport scattering time. In our work, we consider three possible contributions to the finite temperature scattering: (1) Coulomb scattering mediated by charged disorder arising from random quenched impurities, (2) electron-electron scattering, and (3) electron-phonon scattering.

We start by discussing disorder scattering, which is present even at $T=0$, producing a substantial residual resistivity in the data of Ref.~\cite{zhao2022gate}. The Coulomb interaction in 2D is given by $V_{\bq} = \frac{2\pi e^2}{\kappa q}$. The bare Coulomb potential of the charged disorder is screened by the carriers themselves, resulting in the screened potential
\beq
u_{\bq}=\frac{V_{\bq}}{\epsilon(q,T)} = \frac{2\pi e^2}{\epsilon(q,T)\kappa q},\label{eq:coulomb}
\eeq
where, within the RPA screening theory, $\epsilon(q,T)$ is the finite-temperature static RPA screening function in 2D given by 
\beq
\epsilon(q,T) =1+\frac{2\pi e^2}{\kappa q}\Pi(q,T).
\eeq
Here, $\Pi(q,T)$ is the 2D finite-temperature static polarizability function. At $T=0$, the polarizability function is given by~\cite{stern1967polarizability}
\begin{align}
    \Pi(q,T=0)=N_F\left[1-\sqrt{1-(2k_F/q)^2}\Theta(q-2k_F)\right],
\end{align}
where $N_F=gm/2\pi\hbar^2$ is the 2D density of states with $g=2$ the total degeneracy (spin), $k_F$ is the Fermi momentum, and $\Theta(x)$ is the Heaviside step function. At $T=0$, the disorder scattering occurs on the Fermi surface. As a result, the maximum value of $q$ is given by $0\le q\le 2k_F$, with $2k_F$ scattering being the dominant resistive scattering process. Within this regime, the polarizability function is constant and the screening function is $s$-wave. The finite-temperature polarizability function can be calculated using the corresponding $T=0$ function via~\cite{sarma2004metallicity,gold1986temperature,maldague1978many}
\begin{align}
    \Pi(q,T)&= \frac{\beta}{4}\int_0^\infty d\mu' \frac{\Pi(q,T=0)|_{\epsilon_F=\mu'}}{\cosh^2\frac{\beta}{2}\left(\mu-\mu'\right)},
\end{align}
where $\beta=1/k_BT$, $\epsilon_F$ is the Fermi energy, and $\mu$ is the finite-temperature chemical potential given by
\beq
\mu(T)=k_B T\ln\left[\exp\left(\frac{T_F}{T}\right)-1\right].
\eeq
The primary effect of finite temperature is to smoothen the kink of the polarizability function at $q=2k_F$. This thermal smoothening of the $2k_F$ kink has a nonanalytic form in 2D, leading to a strong suppression of screening, and hence to strong temperature dependence of the resistivity arising from impurity scattering in 2D which does not happen in 3D systems. This 2D Fermi surface anomaly leads to linear-in-$T$  Fermi liquid corrections to various properties, including the resistivity, rather than the generic $O(T^2)$ leading order corrections expected from the Sommerfeld expansion. One can think of the anomalous resistive scattering as arising from the strongly temperature dependent 2D Friedel oscillations associated with the impurity screening~\cite{zala2001interaction}. 

The transport scattering time from charged disorder at energy $\epsilon_{\bk}=\hbar^2k^2/2m$ is given by \cite{stern1980calculated}
\begin{align}
    \frac{1}{\tau(\epsilon_{\bk})}= \frac{2\pi n_i}{\hbar} \int_{\kk'} \frac{d^2\kk'}{(2\pi)^2}\left|u_{\kk-\kk'}\right|^2 (1-\cos\theta_{\kk'\kk}) \delta(\epsilon_{\kk'}-\epsilon_\kk),
\end{align}
where  $n_i$ is the charged impurity density, $\theta_{\bk'\bk}$ is the angle between $\bk$ and $\bk'$, and $u_{\bq}$ is the screened Coulomb interaction [see Eq.~(\ref{eq:coulomb})]. At finite temperatures, the scattering time is calculated in the Boltzmann transport theory by averaging over all energies via
\beq
\tau(T) = \frac{\int d\epsilon \epsilon \tau(\epsilon) \lt(
-\frac{\partial f}{\partial \epsilon}\rt)}{\int d\epsilon \epsilon \lt(
-\frac{\partial f}{\partial \epsilon}\rt)},\label{eq:disorder_full}
\eeq
where
$f(\epsilon) = 1/\left[\exp\lt(\frac{\epsilon-\mu}{k_BT}\rt)+1\right]$
is the Fermi distribution function. Using this expression for $\tau(T)$, we can asymptotically expand $\rho(T)$ in the two opposite limits of low temperatures $T\ll T_F$ and high temperatures $T\gg T_F$, which gives~\cite{das2015screening,sarma2003low,sarma2004metallicity}
\begin{align}
    \rho(T\ll T_F) &= \rho_0 \left(1+\frac{2q_s}{q_s+1}\frac{T}{T_F}\right),\label{eq:disorder_lowT}\\
    \rho(T\gg T_F)&= \rho_1\frac{T_F}{T},\label{eq:disorder_highT}
\end{align}
where $q_s=q_{TF}/2k_F$ with $q_{TF}=\frac{gme^2}{\kappa\hbar^2}$ as the 2D Thomas-Fermi screening wavevector, $\rho_0$ is the residual resistivity at $T=0$, and $\rho_1=(h/e^2)(n_i/n)(\pi q_s^2/2)$. We note that at low $T$, the resistivity rises linearly in $T$ from the $T=0$ residual resistivity value, and at high $T$, the impurity induced resistivity decreases as $1/T$, implying an impurity scattering induced  resistivity maximum around $T \sim T_F$.

At very low temperatures, the temperature dependence of the resistivity is additionally suppressed due to the impurity scattering induced broadening of the Fermi surface. The resulting temperature dependence approximately follows~\cite{das2015screening,sarma2014mobility}
\begin{align}
    \rho(T)\approx \rho_0\left[1+a T\exp(-T_D/T)\right],\label{eq:disorder_dingle}
\end{align}
where $a=\frac{2q_s}{q_s+1}$, and $T_D$ is the Dingle temperature defined by $T_D=\hbar/2k_B\tau_q$, where $\tau_q$ is the single-particle relaxation time. 

Next, we discuss electron-electron scattering. This manifestly temperature-dependent scattering mechanism is dominant in a clean Fermi liquid, where the electrons can undergo crystal-momentum preserving Umklapp scattering mediated by the inter-electron Coulomb interaction. This leads to a finite lifetime of the electrons, the exact form of which depends on the details of the material and the microscopic model used. We will assume a scattering rate of the form arising in the standard 2D Fermi liquid theory~\cite{zheng1996coulomb,sarma2021know,ahn2022planckian,behnia2022origin}
\begin{align}
    \frac{1}{\tau(T)}\approx \frac{\pi \epsilon_F}{4\hbar}\left(\frac{k_BT}{\epsilon_F}\right)^2,
\end{align}
which is valid in the limit $T\ll T_F$. This leads to, assuming Umklapp scattering, a temperature-dependent resistivity of the form 
\begin{align}
    \rho(T\ll T_F)=\rho_0 +\frac{\pi^2}{2}\frac{\hbar}{e^2}\left( \frac{T}{T_F}\right)^2,\label{eq:ee_scattering}
\end{align}
where $\rho_0$ is the $T=0$ residual resistivity caused by disorder scattering. Note that the electron-electron scattering by itself cannot produce a $T=0$ residual resistivity, which is clearly observed in Ref.~\cite{zhao2022gate}, indicating the presence of considerable disorder scattering. Writing Eq.~(\ref{eq:ee_scattering}) as $\rho(T)=\rho_0+AT^2$, Ref.~\cite{zhao2022gate} uses the quadratic coefficient $A$ to estimate the effective band mass since $\sqrt{A}\sim 1/T_F \sim m$, assuming a known carrier density.

Finally, we discuss the contribution of phonon scattering to the resistivity, which also operates only at finite temperatures, and is negligible for $T\ll T_{BG}$, where $T_{BG}$ is the Bloch-Gr\"{u}neisen temperature. (We mention that in normal metals, $T_{BG}\gg$ Debye temperature $T_{\rm Debye}$, and hence $T_{\rm Debye}$ is the lower cut off temperature scale for phonon scattering, but in doped semiconductors, such as TMDs, $T_{BG}\ll T_{\rm Debye}$, and hence $T_{BG}$ is the lower cut off scale for phonon scattering.) For the Moir\'{e} systems under consideration, the Bloch-Gr\"{u}neisen temperature follows $T_{BG}\sim 20\sqrt{n/10^{12}\,\mathrm{cm}^{-2}}$~K, where $n$ is in units of $\rm cm^{-2}$~\cite{kaasbjerg2013acoustic}. At temperatures above $T_{BG}$, the phonon contribution to the resistivity is linear in $T$, and is given by~\cite{min2012interplay,hwang2019linear}
\begin{align}
    \rho_{ph}(T\gg T_{BG}) &= \rho_0+A_{ph}{T},\label{eq:phonon}
\end{align}
where the slope $A_{ph}$ is related to the dimensionless parameter $\lambda$ that characterizes the electron-phonon coupling strength~\cite{hwang2019linear}:
\begin{align}
    A_{ph}=\frac{m}{ne^2}\left(2\pi\lambda\right) \frac{k_B}{\hbar},
\end{align}
where $\lambda\sim 0.05$ for TMD materials~\cite{kaasbjerg2013acoustic,kaasbjerg2019electron}. We are using an approximate estimated value of the electron-phonon coupling for TMDs combining both the deformation potential and piezoelectric interactions. Below $T_{BG}$, the phonon contribution to $\rho(T)$ is heavily suppressed and goes as $\sim T^4$:
\beq
\rho_{ph}(T\ll T_{BG}) = \rho_0+B_{ph}\left(\frac{T}{T_{BG}}\right)^4.
\eeq
In our analysis, we will assume the functional form of the phonon resistivity contribution to be
\beq
\rho_{ph}(T) = A_{ph}(T-T_{BG}) \Theta(T-T_{BG}),\label{eq:phonon_empirical}
\eeq
neglecting the phonon contribution below $T_{BG}$, assuming that the corresponding $O(\lt(T/T_{BG}\rt)^4)$ contribution to $\rho_{ph}(T)$ is negligible (which is true).


\section{Low-temperature behavior: $T$ vs $T^2$}
In Ref.~\cite{zhao2022gate}, experimental data for $\rho(T)$ is presented at different filling factors $\nu=1+x$ and $\nu=2+x$ written in terms of the independent filling factors of the two layers $\nu=\nu_{\mathrm{Mo}}+\nu_\mathrm{W}$. Based on the scattering models introduced above, we now provide qualitative fits to the experimental data and discuss the relative importance of the different scattering mechanisms in different temperature regimes. Throughout this paper, we mainly focus on the case $\nu=1+x$ and only briefly comment on the case of $\nu=2+x$ at the end. This is consistent with the emphasis in Ref.~\cite{zhao2022gate}.

\begin{figure}
    \centering
    \includegraphics[width=\columnwidth]{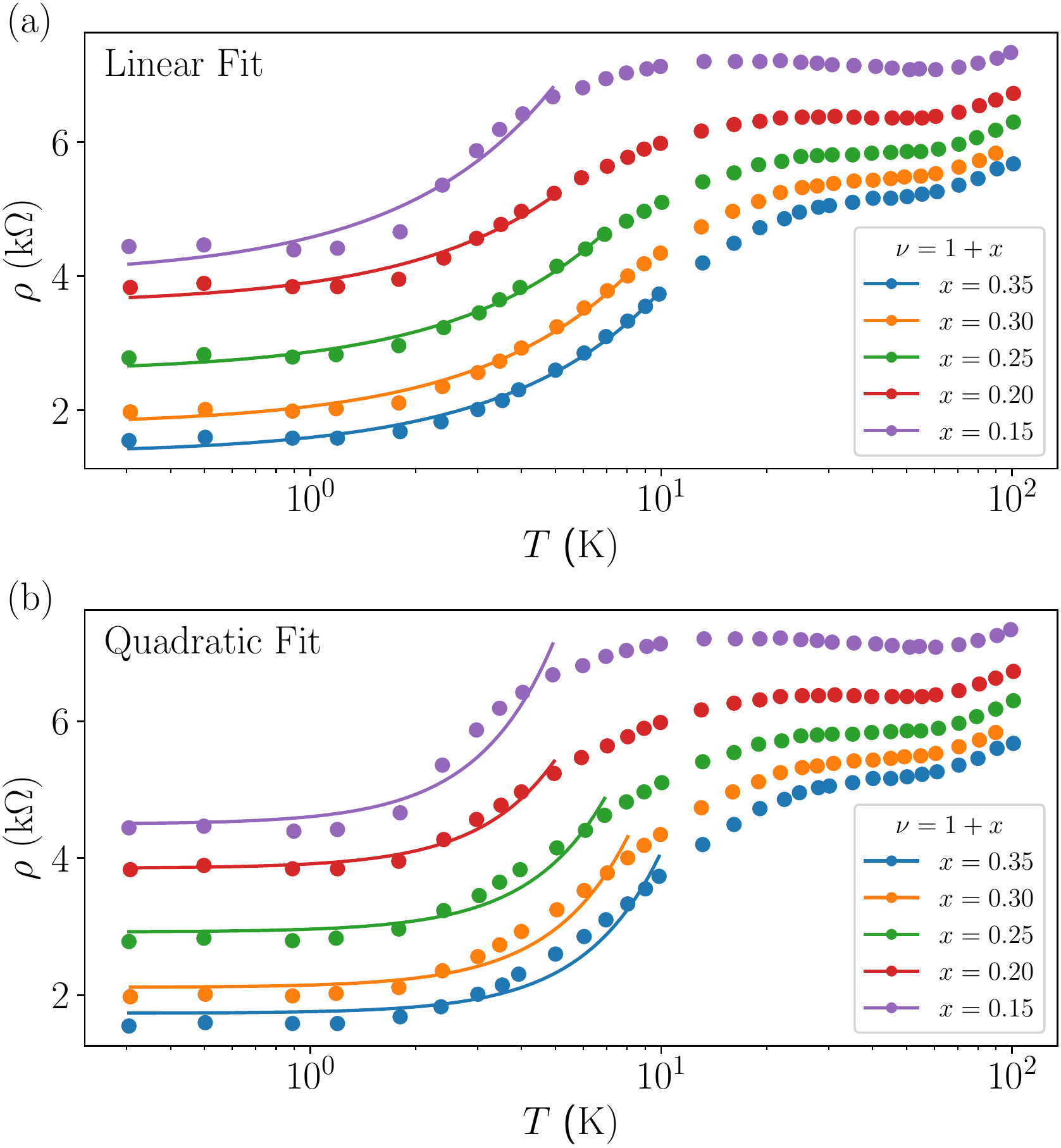}
    \caption{Fits (solid lines) to the low-temperature experimental resistivity data $\rho(T)$ taken from Ref.~\cite{zhao2022gate} (dots) at filling $\nu=1+x$. (a) Linear fit (disorder scattering) and (b) quadratic fit (Umklapp electron-electron scattering). We find that the linear fit is in better agreement with the experimental data as compared to the quadratic fit, signifying the importance of disorder scattering at low temperatures.}
    \label{fig:fitfilling1}
\end{figure}

\begin{figure*}[tb]
    \centering
    \includegraphics[width=\textwidth]{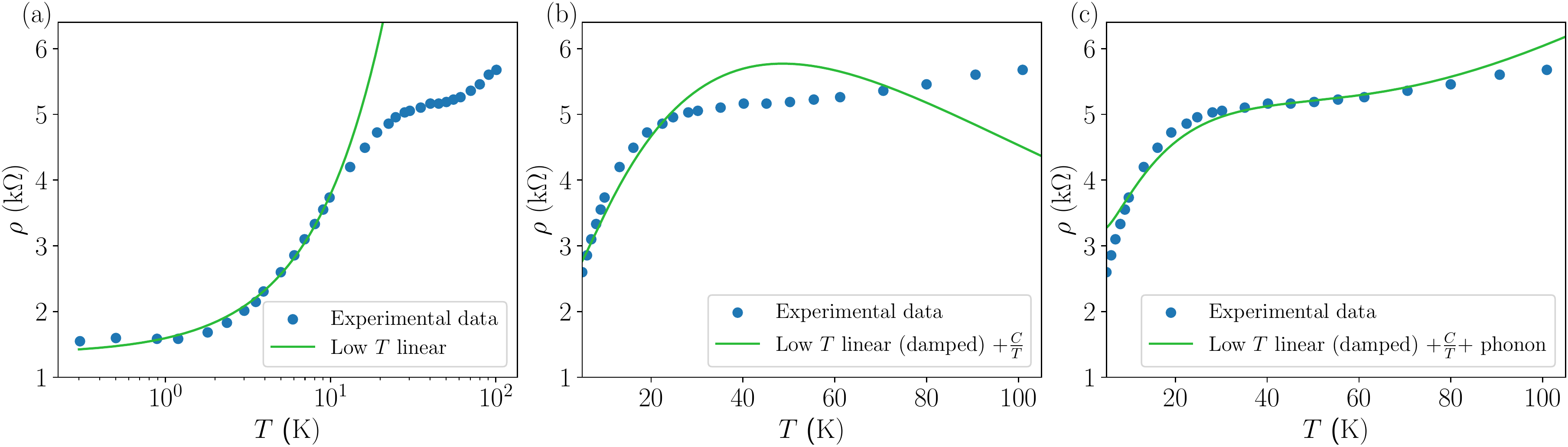}
    \caption{Fits to the experimentally measured resistivity $\rho(T)$ at filling $\nu=1+0.35$, taken from Ref.~\cite{zhao2022gate}, assuming screened disorder scattering and electron-phonon scattering as the primary scattering mechanisms. (a) Linear fit in the low-temperature regime, including only disorder scattering [same as Fig.~\ref{fig:fitfilling1}(a)]. (b) Combined fit capturing the crossover of the disorder contribution from $\rho(T)\sim T$ at $T\ll T_F$ to $\rho(T)\sim 1/T$ at $T\gg T_F$, see Eq.~(\ref{eq:fitfig2b}). (c) Fit including the contribution of disorder scattering to the resistivity [as in panel (b)] with an additional linear-in-$T$ contribution from phonon scattering, see Eq.~(\ref{eq:fitfig2c}), which reproduces the experimental data reasonably well.}
    \label{fig:fithighT}
\end{figure*}

\begin{table}[tb]
\begin{center}
\caption{Fermi temperature $T_F$ and effective mass $m$ as extracted from the linear fit (disorder scattering) to the experimental resistivity data of Ref.~\cite{zhao2022gate} at filling $\nu=1+x$. Values with subscript 1 [subscript 2] are obtained using $n=xn_M$ [$n=(1+x)n_M$].}
\label{tab:disorder_nu1}
\begin{ruledtabular}\begin{tabular}{ccccc} 
$x$ & $T_{F}|_{1}\ (\mathrm{K})$ &  $m/m_e|_{1}$&  $T_{F}|_{2}\ (\mathrm{K})$ &  $m/m_e|_{2}$\\\hline
 $0.35$ & $10.7$ & $4.5$ & $10.9$ & $17.5$ \\
 $0.30$ & $12.4$ & $3.4$ & $12.7$ & $14.6$ \\
 $0.25$ & $16.0$ & $2.2$ & $16.5$ & $10.9$ \\
 $0.20$ & $21.5$ & $1.3$ & $22.7$ & $7.8 $\\
 $0.15$ & $12.6$ & $1.6$ & $13.1$ & $12.7 $\\
 \end{tabular}
\end{ruledtabular}
\end{center}
\end{table}

\begin{table}[tb]
\begin{center}
\caption{Fermi temperature $T_F$ and effective mass $m$ as extracted from the quadratic fit (Umklapp electron-electron scattering) to the experimental resistivity data of Ref.~\cite{zhao2022gate} at filling $\nu=1+x$. Values with subscript 1 [subscript 2] are obtained using $n=xn_M$ [$n=(1+x)n_M$]. Note that in this model, $T_F$ is independent of $n$.}
\label{tab:ee_nu1}
\begin{ruledtabular}\begin{tabular}{ccccc} 
${x}$ & $T_F\ (\mathrm{K})$ &$m/m_e|_1$ &$m/m_e|_2$\\\hline
 $0.35$ & $29.4$ & $1.7$ & $6.3$ \\
 $0.30$ & $26.1$ & $1.6$ & $6.9$ \\
 $0.25$ & $22.3$ & $1.6$ & $7.8$ \\
 $0.20$ & $22.3$ & $1.2$ & $7.4$ \\
 $0.15$ & $11.2$ & $1.9$ & $14.3$ \\
 \end{tabular}
\end{ruledtabular}
\end{center}
\end{table}

The first question we consider is whether the mechanism responsible for the low-temperature variation of resistivity is primarily electron-electron scattering or Coulomb disorder scattering. Note that, by definition, disorder scattering is important at the lowest temperatures since it contributes to the residual resistivity $\rho_0$, with a leading order linear-in-$T$ correction, and the electron-electron scattering vanishes at the lowest temperatures (and also in the absence of any Umklapp scattering), and is quadratic in temperature. The distinguishing factor is clear: For $T\ll T_F$, the electron-electron scattering contribution to the resistivity depends quadratically on temperature, $\rho=\rho_0+A T^2$, with the coefficient $A>0$, while the disorder scattering contribution follows a linear dependence, $\rho=\rho_0(1+B T)$, with the coefficient $B>0$. In Fig.~\ref{fig:fitfilling1}, we directly compare a linear fit [Fig.~\ref{fig:fitfilling1}(a)] with a quadratic fit [Fig.~\ref{fig:fitfilling1}(b)] with respect to the experimentally measured $\rho(T)$ in Ref.~\cite{zhao2022gate} at filling $\nu=1+x$. We find that the linear fit is generally better than the quadratic one, signifying that disorder scattering by itself is a plausible explanation for the observed low-temperature resistivity. Of course, in reality, both scattering mechanisms may be present simultaneously. However, we do not fit a combination of a linear and a quadratic function since we are only interested in studying the \emph{dominant} scattering mechanism and want to keep the number of fitting parameters to a minimum.  (Also, the data of Fig.~\ref{fig:fitfilling1} are simply not accurate enough for theoretical fits to multiple combined power laws in temperature.)

Next, we extract the relevant physical parameters from the best fits shown in Fig.~\ref{fig:fitfilling1}. In the case of disorder scattering, we use Eq.~(\ref{eq:disorder_lowT}) together with the relations $q_s(T_F,m) =\frac{ge^2}{2\kappa\hbar}\sqrt{\frac{m}{2k_BT_F}}$ and $m(T_F)=\frac{2\pi n\hbar^2}{gk_BT_F}$ to relate the linear coefficient $B$ to the Fermi temperature $T_F$, while keeping the density $n$ fixed. We consider two scenarios, $n=xn_M$ and $n=(1+x)n_M$, where $n_M\approx 5\times 10^{12}$~cm$^{-2}$ is the Moir\'{e} density reported in Ref.~\cite{zhao2022gate}. The former scenario is what is expected for a doped Mott insulator, while the latter scenario is the carrier density determined from experimental Hall measurements~\cite{zhao2022gate}. The resulting values for $T_F$ and $m$ are shown in Table~\ref{tab:disorder_nu1}. Overall, we find that these values are consistent with experimental parameters within an order of magnitude. The band mass is larger than the experimental value $m=0.5m_e$, which we attribute to the uncertainties in the experimental carrier density and residual $T=0$ resistivity~\cite{faiprivate}. We note that the band mass is enhanced in the case of $n=(1+x)n_M$ by about a factor of $10-20$ with respect to the $n=xn_M$ case since the carrier density increases by a factor of $10$. In order to keep $T_F$ the same (which is a function of $n/m$), the band mass needs to scale with the carrier density. Furthermore, we have calculated the Anderson localization critical density (for $x<0.35$) defined by the Ioffe-Regel-Mott criterion, $\epsilon_F\tau=1$~\cite{sarma2014two}, which is consistent with unpublished data~\cite{faiprivate} associated with Ref.~\cite{zhao2022gate}. Note that it was already emphasized in Ref.~\cite{ahn2022disorder} that doped homobilayer TMDs undergo a low-doping Anderson-type metal-insulator localization transition induced by Coulomb disorder, and our current work shows the same to be true for the heterobilayer TMDs of Ref.~\cite{zhao2022gate}---both arise from the presence of strong Coulomb disorder in TMD samples with the low-$T$ mobility being only around a few thousand $\rm cm^2/V\cdot s$, indicating the presence of substantial ($\sim 10^{11}\,\rm cm^{-2}$) charged impurities in the environment as reflected in the rather large residual resistivity.

We also extract the physical parameters from the quadratic fit assuming that electron-electron scattering is the primary scattering mechanism. Using Eq.~(\ref{eq:ee_scattering}), we relate the fitting coefficient $A$ to the Fermi temperature $T_F$. The band mass is then found as a function of $T_F$ and $n$, which we again assume to be fixed to either $n=xn_M$ or $n=(1+x)n_M$. Our results are summarized in Table~\ref{tab:ee_nu1}. Again, we find that the Fermi temperatures extracted from the fit are consistent with experimental values, while the band masses are about a factor of $10$ larger than the experimentally reported $m=0.5m_e$.

\section{High-temperature behavior}
We now proceed to study the high-$T$ behavior of $\rho(T)$ using a combination of qualitative fits and exact theory. \figref{fig:fithighT} shows a fit to the full experimental resistivity data at filling $\nu=1+0.35$. Here, we proceed in three steps: First, we perform a linear fit $\rho(T) = \rho_0(1+ BT)$ for $\rho_0$ and $B$ to the low-temperature data, taking disorder scattering to be the dominant scattering mechanism in this regime [\figref{fig:fithighT}(a)]. This is the same fit as the one shown in Fig.~\ref{fig:fitfilling1}(a). Second, at high temperatures, the linear $\rho(T)\sim T$ dependence from disorder scattering undergoes a smooth crossover to $\rho(T)\sim 1/T$, see Eq.~(\ref{eq:disorder_highT}). To capture this crossover, we perform an additional fit to the high-temperature resistivity data using a fitting function of the form 
\begin{align}
\rho(T) = \rho_0[1+B T\exp(-T/T_{0})]+\frac{C}{T},\label{eq:fitfig2b}
\end{align}
with $T_0$ and $C$ as the fitting parameters and where $\rho_0$ and $B$ are fixed to the values obtained from the low-temperature fit discussed above. Here, the factor of $\exp(-T/T_0)$ is a phenomenological way of suppressing the linear-in-$T$ behavior (valid for $T\ll T_F$) and capturing the smooth quantum-classical crossover to $\rho(T)\sim 1/T$ (for $T\gg T_F$). The resulting curve is shown in \figref{fig:fithighT}(b). This is not a very good fit since we need a third ingredient: the linear-in-$T$ phonon scattering that sets in at high temperatures $T>T_{BG}$. Therefore, we add a term given by $A_{ph}T$, with $\lambda=0.05$, and recalculate the fit for $C$ and $T_0$ using the functional form
\begin{align}
\rho(T) = \rho_0[1+B T\exp(-T/T_{0})]+\frac{C}{T}+A_{ph}T,\label{eq:fitfig2c}
\end{align}
to obtain the curve in \figref{fig:fithighT}(c), which is now in good agreement with the experimental data. Thus, disorder scattering (including quantum-classical crossover and Friedel oscillation effects) plus phonon scattering provides an acceptable explanation for the temperature-dependent resistivity in Ref.~\cite{zhao2022gate} without the necessity for invoking any Umklapp electron-electron scattering.

\begin{figure}[tb]
    \centering
    \includegraphics[width=\columnwidth]{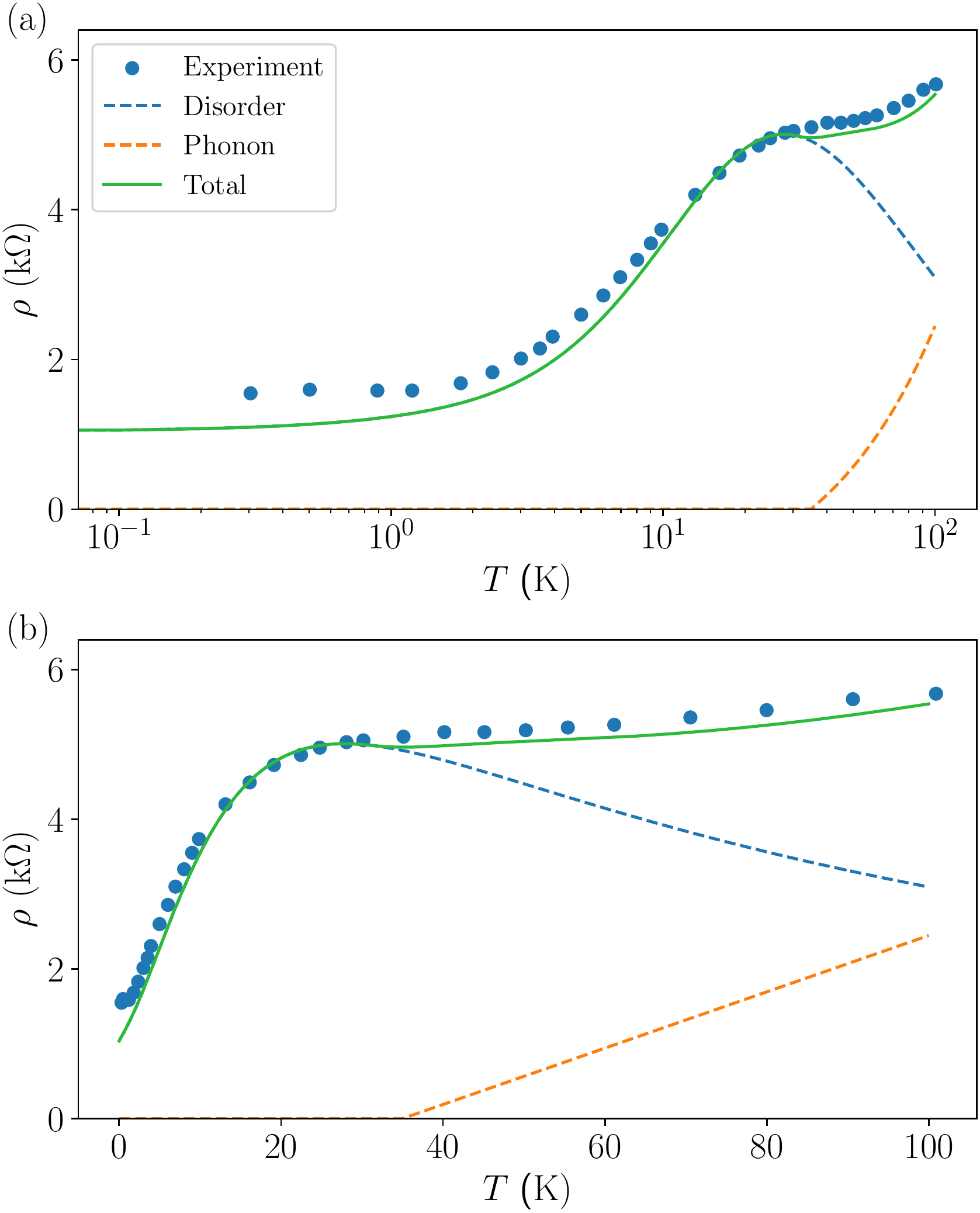}
    \caption{Contribution to $\rho(T)$ from disorder scattering calculated numerically using Eq.~(\ref{eq:disorder_full}) (dashed blue line), empirical phonon resistivity as given in Eq.~(\ref{eq:phonon_empirical}) (dashed orange line), total resistivity including both disorder scattering and phonon scattering (green line), compared with the experimental resistivity data taken from Ref.~\cite{zhao2022gate} at filling $\nu=1+0.35$ (blue dots). The parameters used are $n_i=6\times 10^{10}$~cm$^{-2}$, $T_F=11$~K, $T_{BG}=35$~K, and $\lambda=0.05$. (a) and (b) show the same results, but on a logarithmic temperature scale and a linear temperature scale respectively.}
    \label{fig:fullTdisorder}
\end{figure}

\begin{table}[tb]
\begin{center}
\caption{Fermi temperature $T_F$, effective mass $m$, and Dingle temperature $T_D$ as extracted from the linear fit (disorder scattering) to the experimentally reported resistivity of Ref.~\cite{zhao2022gate} at filling $\nu=2+x$, using $n=xn_M$.}
\label{tab:disorder_nu2}
\label{fitdisorder1plusx}\begin{ruledtabular}\begin{tabular}{cccc} 
$x$ & $T_{F}\ (\mathrm{K})$ &  $m/m_e$ & $T_D\ (\mathrm{K})$\\\hline
 $0.35$ & $30.5$ & $1.6$ & $36.0$ \\
 $0.30$ & $32.0$ & $1.3$ & $38.1$ \\
 $0.25$ & $35.0$ & $1.0$ & $37.8$ \\
 $0.20$ & $41.9$ & $0.7$ & $36.4$ \\
 $0.15$ & $49.5$ & $0.4$ & $43.8$ \\
 \end{tabular}
\end{ruledtabular}
\end{center}
\end{table}

For completeness, we now directly calculate $\rho(T)$ treating the disorder scattering exactly at all temperatures by numerically evaluating Eq.~(\ref{eq:disorder_full}) and only adding the phonon contribution later by hand, see Eq.~(\ref{eq:phonon}). The minimal parameters needed to perform this calculation are the impurity density $n_i$, the Fermi temperature $T_F$, the Bloch-Gr\"{u}neisen temperature $T_{BG}$, and the electron-phonon coupling $\lambda$. Based on qualitative estimates, we choose $n_i=6\times 10^{10}\,{\rm cm^{-2}}, T_F=11\,{\rm K}, T_{BG}=35\,{\rm K}$ and $\lambda=0.05$. We emphasize that these parameters are not obtained by fitting but are estimated based on our analysis of the experimental low-$T$ data (including fitting $\rho_0$) in the previous section. The resulting numerical $\rho(T)$ is shown in \figref{fig:fullTdisorder}(a) using a logarithmic temperature scale, emphasizing the comparison with the experimental data at low $T$, as Ref.~\cite{zhao2022gate} presents the results on a logarithmic scale. We see that the numerically calculated $\rho(T)$ is consistent with the experimental data. \figref{fig:fullTdisorder}(b) shows the same curve in a linear temperature scale, emphasizing the comparison with the experimental data at high $T$. Here, the numerically calculated $\rho(T)$ is in very good agreement with the experimental data. We emphasize again that our analysis is based on qualitative estimates of the relevant system parameters and no quantitative predictions can be made at this point because of large uncertainties in the experimental data of Ref.~\cite{zhao2022gate}\cite{wenjinprivate}. Nevertheless, our result clearly shows that the combination of disorder and phonon scattering is a plausible explanation for the temperature dependence of the resistivity measured in Ref.~\cite{zhao2022gate}.


\begin{figure}[tb]
    \centering
    \includegraphics[width=\columnwidth]{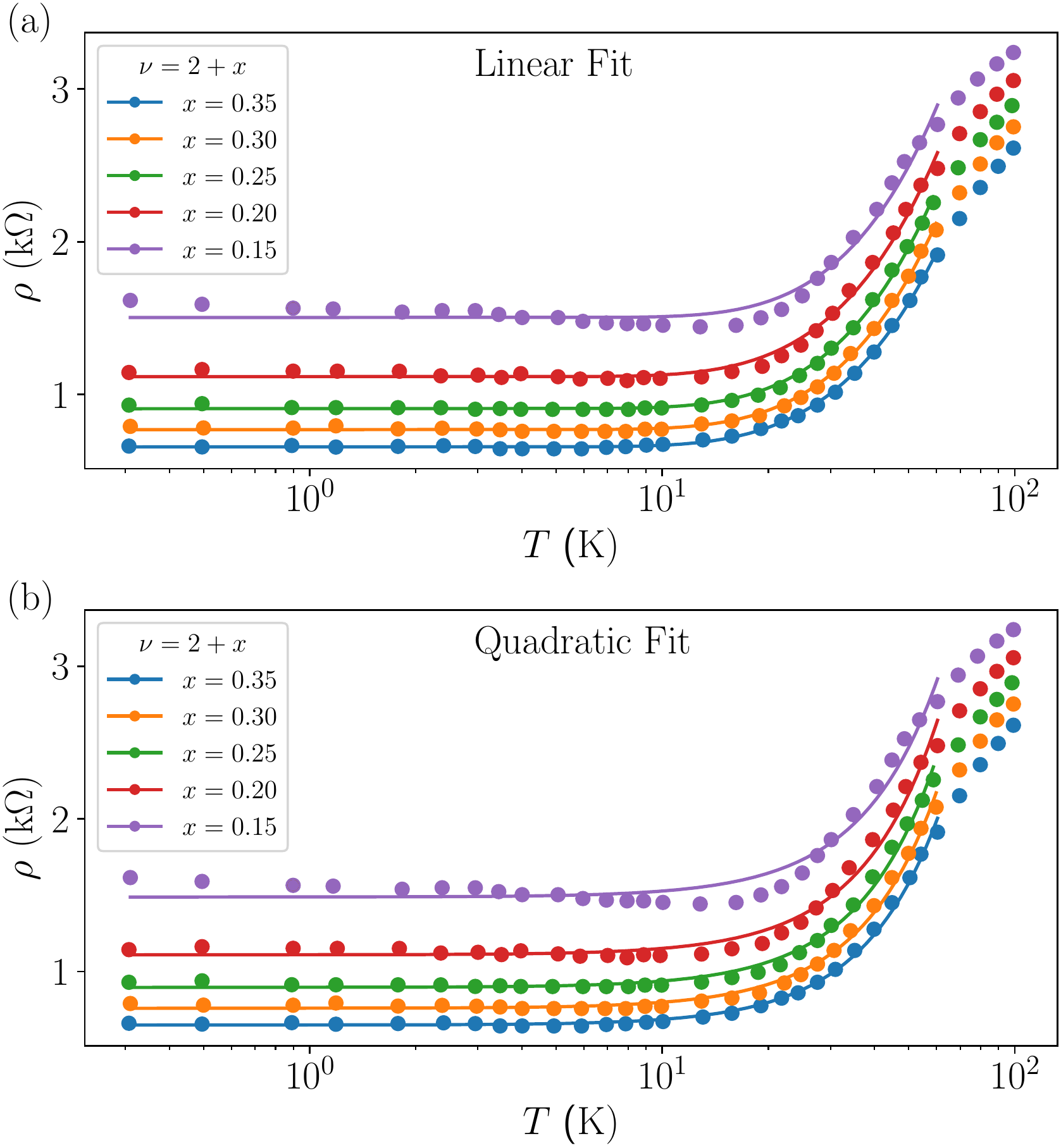}
    \caption{Fits (solid lines) to the low-temperature experimental resistivity data $\rho(T)$ taken from Ref.~\cite{zhao2022gate} (dots) at filling $\nu=2+x$. (a) Linear fit (disorder scattering) and (b) quadratic fit (Umklapp electron-electron scattering). Both fits agree reasonably well with the experimental data.}
    \label{fig:fitmakdata3}
\end{figure}

\begin{table}[tb]
\begin{center}
\caption{Fermi temperature $T_F$ and effective mass $m$ as extracted from the quadratic fit (Umklapp electron-electron scattering) to the experimentally reported resistivity data of Ref.~\cite{zhao2022gate} at filling $\nu=2+x$, using $n=xn_M$.}
\label{tab:ee_nu2}
\label{fitdisorder1plusx}\begin{ruledtabular}\begin{tabular}{ccc} 
$x$ & $T_{F}\ (\mathrm{K})$ &  $m/m_e$ \\\hline
 $0.35$ & $229.1$ & $0.21$ \\
 $0.30$ & $220.2$ & $0.19$ \\
 $0.25$ & $212.6$ & $0.16$ \\
 $0.20$ & $209.3$ & $0.13$ \\
 $0.15$ & $220.5$ & $0.09$ \\
 \end{tabular}
\end{ruledtabular}
\end{center}
\end{table}

\section{Filling factor $\nu=2+x$}
For completeness, we also comment on the low-$T$ behavior of $\rho(T)$ at fillings $\nu=2+x$. Again, we perform both a linear and a quadratic fit for $\rho(T)$ at low temperatures in order to compare the importance of electron-electron scattering and disorder scattering. One of the key differences between the resistivity-vs-temperature curves for $\nu=1+x$ and $\nu=2+x$ is that the latter exhibit an extended low-temperature region where the resistivity is only very weakly temperature-dependent. This necessitates the inclusion of a non-zero Dingle temperature in the linear temperature dependence resulting from disorder scattering, see Eq.~(\ref{eq:disorder_dingle}). In \figref{fig:fitmakdata3}, we compare a linear fit including a finite Dingle temperature [\figref{fig:fitmakdata3}(a)] and a quadratic fit [\figref{fig:fitmakdata3}(b)] to the experimental resistivity data. Unlike for $\nu=1+x$, the quadratic and linear fits perform similarly and are in fact hard to distinguish by eye. However, extracting physical quantities from the fit parameters reveals stark differences between the two cases. The linear fit (disorder scattering) results in values of $T_F$, $T_D$, and $m$ that are consistent with experimental estimates (see Table~\ref{tab:disorder_nu2}). The quadratic fit (electron-electron scattering), on the other hand, gives $T_F\sim 200$~K  (see Table~\ref{tab:ee_nu2}), which is more than an order of magnitude larger than the experimentally expected $T_F\sim 10$~K.

\section{Conclusions}
We have critically analyzed the temperature dependence of the resistivity presented in Ref.~\cite{zhao2022gate}, which claims the realization of a Kondo lattice in a Moir\'{e} TMD bilayer system. While Ref.~\cite{zhao2022gate} infers the emergence of heavy fermions from the low-temperature resistivity data assuming that electron-electron Fermi liquid Umklapp scattering as the dominant scattering mechanism, our unbiased comparison of theoretical fits to the experimental data shows that disorder scattering is certainly a viable (if not better) explanation for the resistivity data at low temperatures. Furthermore, at higher temperatures, including a contribution from phonon scattering along with the disorder scattering, allows us to reproduce the experimentally observed resistivity plateau. Our work shows that a combination of disorder and phonon scattering provides a reasonable explanation for the resistivity data reported in Ref.~\cite{zhao2022gate} and that additional experimental signatures are needed to reach a decisive conclusion as to whether a Kondo lattice has been observed or not. Given that Ref.~\cite{zhao2022gate} reports substantial levels of the residual resistivity, impurity scattering is manifestly important, and 2D Fermi surface anomalies automatically lead to a strong temperature dependence in the resistivity arising from the scattering by the Friedel oscillations associated with the quenched charged impurities in the system producing the residual resistivity.

\section*{Acknowledgments} %
We thank K. F. Mak and W. Zhao for helpful discussions on the experimental data and for providing us with unpublished results. This work is supported by the Laboratory for Physical Sciences through the Condensed Matter Theory Center.

\bibliography{refs.bib}
\end{document}